\documentclass[sigconf]{acmart}
\AtBeginDocument{%
  \providecommand\BibTeX{{%
    \normalfont B\kern-0.5em{\scshape i\kern-0.25em b}\kern-0.8em\TeX}}}

\usepackage{longtable}
\usepackage{framed}
\usepackage[subtle]{savetrees}

\copyrightyear{2024}
\acmYear{2024}
\setcopyright{rightsretained}
\acmConference[EASE 2024]{28th International Conference on Evaluation and Assessment in Software Engineering}{June 18--21, 2024}{Salerno, Italy}
\acmBooktitle{28th International Conference on Evaluation and Assessment in Software Engineering (EASE 2024), June 18--21, 2024, Salerno, Italy}\acmDOI{10.1145/3661167.3661229         }
\acmISBN{979-8-4007-1701-7/         24/06}
\begin{document}

\title{What You Use is What You Get: Unforced Errors in Studying Cultural Aspects in Agile Software Development}


\author{Michael Neumann}
\orcid{0000-0002-4220-9641}
\affiliation{%
  \institution{Hochschule Hannover}
  \city{Hannover}
  \country{Germany}}
\email{michael.neumann@hs-hannover.de}

\author{Klaus Schmid}
\orcid{0000-0002-4147-3942}
\affiliation{%
  \institution{University of Hildesheim}
  \city{Hildesheim}
  \country{Germany}}
\email{schmid@sse.uni-hildesheim.de}

\author{Lars Baumann}
\affiliation{%
  \institution{Hochschule Hannover}
  \city{Hannover}
  \country{Germany}}
\email{lars.baumann@hs-hannover.de}

\renewcommand{\shortauthors}{Neumann et al.}

\begin{abstract}
\textit{Context:} Cultural aspects are of high importance as they guide people's behaviour and thus, influence how people apply methods and act in projects. In recent years, software engineering research emphasized the need to analyze the challenges of specific cultural characteristics. Investigating the influence of cultural characteristics is challenging due to the multi-faceted concept of culture. People's behaviour, their beliefs and underlying values are shaped by different layers of culture, e.g., regions, organizations, or groups. In this study, we focus on agile methods, which are agile approaches that focus on underlying values, collaboration and communication. Thus, cultural and social aspects are of high importance for their successful use in practice. 
\textit{Objective:} In this paper, we address  challenges that arise when using the model of cultural dimensions by Hofstede to characterize specific cultural values. This model is often used when discussing cultural influences in software engineering.
\textit{Method:} As a basis, we conducted an exploratory, multiple case study, consisting of two cases in Japan and two in Germany.
\textit{Contributions:} In this study, we observed that cultural characteristics of the participants differed significantly from cultural characteristics that would typically be expected for people from the respective country. This drives our conclusion that  for studies in empirical software engineering that address cultural factors, a case-specific analysis of the characteristics is needed.
\end{abstract}





\maketitle

\section{Introduction}
\label{Sec1:Intro}
Software development (SD) is a human-centred discipline and studying the behaviour of SD professionals shaped by their culture has gained an increasing attention in software engineering research~\cite{Amrit.2014}. Especially, the question how cultural aspects influence software development has been intensively studied~\cite{Deshpande.2010,Jaanu.2012,Iivari.2011,Rabelo.2015,Sharp.1999,Sharp.2005}. We know that dealing with the influences of culture is complex as the  concept of culture is multi-faceted and may be identified relative to regions, organizations, groups (like teams or communities) as well as individuals~\cite{Carmel.2001}. Also, these influences interact with specific software development contexts like offshoring, project characteristics, or company structures~\cite{Deshpande.2010}. 

Based on Hofstede's definition:~\cite{Hofstede.1981}: \textit{``The collective programming of the mind that distinguishes the members of one group or category of people from others.''} we regard culture as the basis for shared values, beliefs, and behaviours of a group or society. Thus, culture guides us in terms of ``what is an expected behaviour'' or ``what is implicitly allowed'' and 
``what is forbidden''. Several authors presented descriptive and comparative models to characterize culture in some ways (e.g., \cite{House.2002,HofstedeMinkov.2013,Karahanna.2006,Trompenaars.2012}). In this study, we focus on the well-known model by Hofstede~\cite{Hofstede.2001} as it is used extensively in the area of empirical software engineering (e.g., ~\cite{Alsanoosy.2019,Ayed.2017,Ghinea.2011}).

In the past two decades, agile methods have become state of the art approaches in software development~\cite{VersionOne.2022}. They are based on values and principles, which are defined in the agile manifesto~\cite{Beck.2000}. Their application is defined by guidelines of specific agile methods like Scrum~\cite{Schwaber.2020}. Further, we know that the interplay between technical agility (doing agile) and cultural agility (being agile) is of high importance for a successful use of agile methods (e.g., ~\cite{Diebold.2015,Gregory.2019,Kuepper.2017,Kuchel.2023}). Today, agile methods are used by both, co-located and (partially) distributed software development teams~\cite{Smite.2020}, often spanning different cultural contexts. According to several studies~\cite{Neumann.2023,Strode.2022,Smite.2021}, cultural aspects are one of the major challenges in agile software development. Several authors investigated these challenges in recent years (e.g., ~\cite{Biddle.2018,Gupta.2019,Hoda.2017,Silveira.2019,Smite.2020}). 
In this study, we focus on agile software development due to the underlying value-based approach, the strong focus of social aspects and, thus, the high importance of cultural influences for their successful use in practice~\cite{Chow.2008}. 

The above motivates the objective of our paper:
Based on our collected data from a multiple case study in Japan and Germany, we first compare our results with Hofstede's data in order to identify differences in the cultural profiles. Next, we discuss existing research using the model by Hofstede in the field of agile software development to identify the potential impact on future research. Based on the identified implications, we finally argue the need of a context-specific assessment of  cultural values in empirical software engineering research.

This paper is structured as follows:  Section~\ref{Sec2Background} gives a brief introduction of Hofstede's model. Section~\ref{Sec3:Research Design} describes our research design, including an overview of the case companies, the data collection, and the analysis approach. 
In Section~\ref{Sec4:Results}, we present the results of our study and a discussion of  recommendations for other researchers. The paper closes with a summary in Section~\ref{Sec6:Conclusion}.

\section{Cultural Dimensions Model by Hofstede}
\label{Sec2Background}
Hofstede's model was initially defined to characterize culture in relation to nations. It consists of six (cultural) dimensions~\cite{Hofstede.2001}, which describe a spectrum of potential cultures:

\textit{Power Distance Index (PDI):} This represents the extent to which individuals with less power accept or expect an unequal distribution of power. An example of this is the behavior towards superiors in a company. Societies with a high PDI value believe in the hierarchy and take  orders without questioning them. Societies with a low PDI value  strive towards equality in terms of the distribution of power.

\textit{Uncertainty Avoidance (UAI):} This dimension describes the degree to which  members of a culture feel comfortable or uncomfortable in new or unknown situations. 
If there is a high degree of uncertainty avoidance, national culture is characterized by clear regulations (such as laws or security measures) and tries to create structures that are as clear as possible. In national cultures that show a low level of uncertainty avoidance, there is a tolerance for other opinions. Besides, regulations are less precise and strict.

\textit{Individualism vs.\ Collectivism (IDV):} According to Hofstede, this describes the degree to which national cultures are  individualistic or collectivist. A low scale value is interpreted as collectivist, a high scale value as individualistic. Members of individualistic societies are interested in their own resources. They primarily take care of themselves and their family. Collectivist societies share resources and a common understanding of moral standards. 

\textit{Masculinity vs.\ Femininity (MAS):} This dimension describes values based on stereotypes commonly associated with masculinity and femininity.  High values indicate a masculine culture, i.e., achievement of success, power, or performance are highly valued. Feminine societies (low MAS value) emphasize relationships and collaboration. This dimension correlates with the IDV dimension.

\textit{Long Term Orientation (LTO):} This describes the degree to which people in a national culture focus on long-term benefits over short-term benefits. 
Long-term cultures can be linked to attributes such as thrift or perseverance. In short-term cultures, the focus is on short-term results to the detriment of long-term effects. This can be seen, for example, in the alignment of control mechanisms in companies and the assessment of executives concerning the short-term aspect.

\textit{Indulgence vs.\ Restraint (IVR):} This dimension indicates the extent to which a national culture enables a free or suppressed satisfaction of human drives (in the context of enjoying life). Also, societies with a focus on indulgence value their free time and do not concentrate on material aspects like the salary. 

For  each cultural value a questionnaires is defined that operationalizes its measurement . Hofstede collected data at IBM in more than 70 countries and used the largest 40 data sets for his initial creation of the comparative model~\cite{Hofstede.2001}. In the past, data collection (also detached from IBM) was continued steadily, so that data for other countries is now available~\cite{Hofstede.2010}. Several sources are available for accessing the data provided by Hofstede~\cite{Hofstede.2001,HofstedeWebsite.2010,HofstedeInsights.2024}.

The results of these measurements are given on a scale from 0..100 with 50 as mid-level. If a score is above 50, the cultural value indicates a relatively high score. Otherwise, if a score is under 50 the value indicates a relatively low score. According to Hofstede~\cite{Hofstede.2001}, all values are relative to describe differences of societies based on a comparison between them. For example, a high measurement value for Individualism  describes a society that is shaped by a culture where everyone prioritizes their personal goals. 

\begin{table*}
 \caption{Overview of the cases and data collection timeline}\vspace*{-1em}
  \label{tab2:OverviewofCases}
   \begin{tabular}{p{0.037\linewidth}p{0.1\linewidth}p{0.52\linewidth}p{0.16\linewidth}p{0.07\linewidth}}
\hline
\textbf{Case ID} & 
\textbf{Company (anonymized)} & 
\textbf{Company profile} &  
\textbf{Unit(s) of Analysis (ASD teams)} & 
\textbf{Country} \\
\hline
C1 & Jap-Comp A & A medium-sized company offering nation-wide contract-based software development (e.g., web sites, online shops, medical products) with ca.\ 50 employees at their software development site. & Three ASD teams with 22 members in total & Japan \\
\hline
C2 & Jap-Comp B & A small-sized  company offering nation-wide  contract-based software development (e.g., XR applications) with ca.\ 15 employees. & One ASD teams with 11 members in total & Japan \\
\hline
C3 & Ger-Comp A & A small-sized company offering nation-wide  contract-based software development (e.g., mobile and web applications) with ca.\ 35 employees. & Three ASD teams with 24 members in total & Germany \\
\hline
C4 & Ger-Comp B & An online marketing company with approx. 1000 employees operating worldwide.
The company develops software for their own needs at five software development sites (e.g., Germany, Poland, UK, USA). Only one site in Germany was studied. & Two ASD teams operating at one site with 17 members in total & Germany \\
\hline
\end{tabular}
\end{table*}

The model of cultural dimensions by Hofstede was discussed and criticized in the past decades (e.g., \cite{Baskerville.2003,Schmitz.2014}). 
Major critical arguments relate to the limitation to a few dimensions, that the national level of culture may not be the best unit for cultural characterization~\cite{McSweeney.2002}, and that it raises the risk of stereotyping~\cite{Carmel.2001}. 
Also, the chosen research design was criticized as surveys may not be an appropriate approach for cultural studies and the representativeness of the population~\cite{Alsanoosy.2020}. Nevertheless, the model is based on a strong empirical basis including a large data set and it has been validated by other researchers. Furthermore, the model applies to several different contexts and has been adopted by various studies (e.g, \cite{Baskerville.2003}). 

\section{Research Design}
\label{Sec3:Research Design}
We conducted an exploratory multiple case study based on the guidelines by Runeson and Höst~\cite{Runeson.2009}. 
Our study is designed as an embedded case study with a defined context as agile software development in Japan and Germany. The cases are  the companies and the units of analysis are the software development teams (see Section~\ref{Sec3-1RContext}). The data collection analysis was performed using the standardized questionnaire and its guideline~\cite{Cultural_Survey} (see Section~\ref{Sec3-2DataAnalysis}). 

\subsection{Research Context}
\label{Sec3-1RContext}
The context of our study is agile software development in Japan and Germany. In total, we studied four companies. 
Table~\ref{tab2:OverviewofCases} provides an overview of the included cases. Due to confidentiality reasons, we had to anonymize the names of the companies as well as the teams and interviewees who participated in our study.

The companies supported us in selecting the software development teams. Per software development site at least one team was studied as a representative of the company/site.

\subsection{Data Collection and Analysis}
\label{Sec3-2DataAnalysis}
The multiple case study was designed to identify cultural influences on agile methods and thus, covered different data collection methods including semi-structured interviews, informal talks, observations of meetings (especially agile practices) and screening of software development related documents. However, in accordance with the objective of our study, we focus in this paper only on determining the cultural values related to the selected model by Hofstede (see Section~\ref{Sec1:Intro}). Thus, we explain below the data collection to determine the cultural values of the agile software development team members using the existing ITIM questionnaire~\cite{Cultural_Survey} as well as the guideline to calculate the cultural values. 

We used the tool SurveyMonkey to collect the data in all cases. For all countries the standardized ITIM questionnaire was translated to the foreign language to reduce language barriers. The translation was performed by native speakers. For the Japanese cases, we had support from the companies, while the German translation was performed by the first author of this paper. In all cases, the questionnaire was open  for three weeks.
As the questionnaire consists of questions related to personal behaviour, beliefs, values, or opinions, it was of high importance to inform about the common quality criteria of empirical research such as transparency, confidentiality, and anonymity.
We had support from the companies to inform about our planned data collection activities. In the German cases, we participated in team meetings to address the survey, inform about the objective of the study and provide our contact details to discuss potential questions and concerns. The Japanese companies supported us in terms of providing mail-addresses of the team members and in translating the information letter. In all four cases, we  sent reminder letters via e-mail after two weeks of data collection. 

The link to the questionnaire was provided to the members of the agile software development teams, thus the population is all members in the teams (see column 4 in Table~\ref{tab2:OverviewofCases}). Overall, the completion rate of the questionnaire is high in all cases: Case 1 (95\%; 21 team members answered), case 2 (100\%; 11 team members answered), case 3 (92\%; 20 team members answered) and case 4 (94\%, 16 team members answered). 

The data analysis, in terms of calculating the cultural values, was performed using the guideline provided by ITIM~\cite{Cultural_Survey} in a Microsoft Excel Sheet for each case. This is available on Zenodo~\cite{Dataset_CulturalProfiles}.

\section{Results}
\label{Sec4:Results}
In this section, we first give an overview of the identified cultural values and compare our results with the data provided by Hofstede. Next, we point to the organizational cultural characteristics and, finally, discuss our results in relation to existing literature as a basis for providing recommendations for future research. 

\subsection{Variation of Cultural Values}
We compared our results with the data, which Hofstede provided in~\cite{HofstedeInsights.2024}, to determine whether there are significant differences between the cultural values that Hofstede determined on a national level, with the specific values we observed within the individual organizations.  
We also verified the data from~\cite{HofstedeInsights.2024} with other references from Hofstede~\cite{Hofstede.2001, HofstedeWebsite.2010} in order to determine whether the values of the two countries are consistent across the publications. Based on the comparison of the data provided by Hofstede, we could identify some differences between Hofstede Insights and other analyses~\cite{Hofstede.2001,HofstedeWebsite.2010}. 
The differences affect the LTO characteristic for Germany 
and the IDV characteristic for both Germany 
and Japan 
However, we decided to use the latest published values as we assume they have the fewest errors.

\begin{figure*}[!tb]
\centering
\includegraphics[scale=0.367]{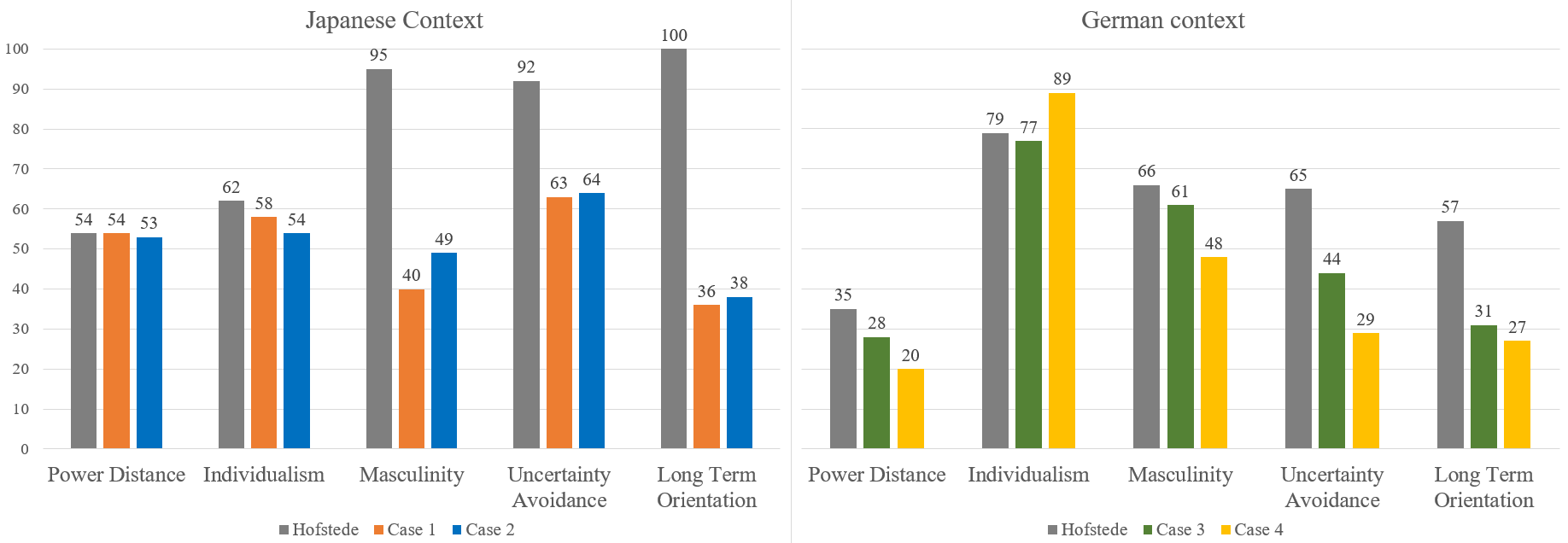}\vspace*{-1em}
\caption{Overview of the compared values}\vspace*{-1em}\label{fig1:ComparisonValues}
\end{figure*}

Figure~\ref{fig1:ComparisonValues} provides a comparison of our calculated cultural values for the German cases and the Japanese cases in relation to the (updated) values provided by Hofstede~\cite{HofstedeInsights.2024}. 

\textit{Power Distance Index (PDI):} In accordance with Hofstede, the values of power distance are low (in terms of less than 50) in both German cases. Further,  case 4 only reaches a score of 20, which indicates a very low power distance. The Japanese cases are close to the Hofstede scores (case 1= 54; case 2= 53 - see Figure~\ref{fig1:ComparisonValues}).

\textit{Individualism vs.\ Collectivism (IDV):} The individualism related cultural characteristics in the German cases are higher than the Hofstede score, especially case 4 with a score of 89 can be interpreted as a highly individualistic society. Both Japanese cases reach slightly higher scores than the mid-level which indicates an individualist characteristic. These values are close to the Hofstede score of 62.

\textit{Masculinity vs.\ Femininity (MAS):} The data from Hofstede scores for the Japanese society a very high masculine characteristic with one of the highest values with 95 in Hofstede's data set. According to Hofstede, the Japanese culture emphasizes the willingness to be the best. In contrast, our data from both cases present scores under mid-level (case 1: 40; case 2: 49) and thus, relate to feminine oriented societies in which the members emphasize quality of life, relationships as well as collaboration. According to Hofstede this cultural dimension correlates to Individualism vs.\ Collectivism. In contrast to the correlation between both dimensions and, e.g., the emphasis of group competitions, our data indicates that collaboration and relationships with other groups are valued. These lower values are clearly in accordance with work in agile environments. Hence, this result may be due to selection effects.

The scores of the German cases are different. In accordance with the data from Hofstede (66), case 3 (61) indicates a masculine society. Related to the Individualism vs.\ Collectivism dimension, 
Case 4 (48) emphasizes feminine characteristics, even if the value is slightly under mid-level. This value is in accordance with the Japanese cases (see the explanation of the correlation to Individualism vs.\ Collectivism above).
Case 3 shows a higher value, which is  characterized as individualist.
However, this tendency is weak and both cases have a lower score than the country score of 66. The value for Case 4 is also very similar to the Japanese situation.

\textit{Uncertainty Avoidance (UAI):} Hofstede's score for Germany (65) shows a preference for uncertainty avoidance in the German culture, emphasizing a focus on regulations and structures. Comparing to our results we identify a difference to Hofstede's score as we determine a significantly lower score: Case 3: 44; Case 4: 29). These low values manifest a willingness for trying new ideas, accept different opinions among the team members and accept the need for new products. 
All these characteristics are in accordance to agile methods, e.g., in terms of continuous optimization, discussing challenges with colleagues and facing challenges when they occur. Thus, this difference can again be explained by the inherent differences between an agile team and the broader population, which Hofstede characterizes. 
Similarly, the Japanese cases (Case 1: 63; Case 2: 64) scores lower significantly lower than Hofstede's data (92). However, as both cases score above mid-level, this cultural characteristic shows a preference for uncertainty avoidance.

\textit{Long Term Orientation (LTO):} We identified a significant difference between all four cases and the data provided by Hofstede. Hofstede scores German (57) and Japanese (88) society as normative, which refers to a focus on long-term benefits. In contrast, our data show low scores for all cases (Case 1: 36; Case 2: 38; Case 3: 31; Case 4: 27). These low values indicate a preference on short-term results and benefits, which is emphasized in agile methods by their iterative-incremental approach.  Again, this is a very significant difference, which is fully in accordance with the difference expected for an agile team in comparison to the broader population.

In summary, our determined cultural values show several differences to the data provided by Hofstede: Our determined cultural values for all cases in both countries differ significantly from the scores provided by Hofstede for some dimensions. Especially related to the dimensions uncertainty avoidance and long term orientation in Germany and masculinity vs. femininity, uncertainty avoidance, and long term orientation in Japan.

\subsection{Implications for Research on Culture}
Based on the comparison of our determined cultural values to the Hofstede scores, we identified several implications for studying cultural influences in empirical software engineering, and in agile software development, in particular:

(1) Our case study results show significant deviations from the data provided by Hofstede. 
These deviations may be related to our  context of agile development. Thus, the subject of the research may strongly influence the cultural values we observe. 
Thus, our findings imply that conducting studies in specific contexts with the aim to analyze cultural aspects solely using the data by Hofstede may lead to strong limitations, 
and   potentially to incorrect  results.

(2) We believe that our case study results also emphasize the added value (if not need) of analyzing the cultural values in the specific context as this allows for a more appropriate assessment of any cultural influences that may exist. 

We emphasize the results above as this is significantly different from the situation in a number of existing research works. 
Below, we discuss the results from existing literature and explain why these studies may be problematic using the data provided by Hofstede. Finally, we present recommendations for the research community especially pointing to the high importance of determining individual cultural values with regard to the context under study. 

To identify a potential for impact on future work in empirical software engineering research, we searched for primary studies using the model as well as the data provided by Hofstede. We found that many studies in computer science and, in particular, in software engineering use the model by Hofstede and the corresponding data, e.g., \cite{Alsanoosy.2018,Alsanoosy.2018a,Borchers.2003,Brockmann.2009,Fischer.2021,Ghinea.2011,Guzman.2018,Hwang.2012,Maris.2007}. We could also identify two  studies dealing specifically with agile software development~\cite{Ayed.2017,Gelmis.2022}. 
The authors of both studies discuss the critical aspects of the model and provide a line of argument for why they decided to use it. Nonetheless, the discussion in both studies lacks in terms of the used data provided by Hofstede. 
Based on our results, it seems questionable whether the cultural values of the participants are similar or even close to Hofstede's data.

Ayed et al.\ use Hofstede's model and data to identify specific influences on agile challenges based on their results from a qualitative study in three countries \cite{Ayed.2017}. 
Further, they identified challenges related to agile software development based on an analysis from the literature and their study results. Their key contribution  are 16 hypothetical correlations between cultural characteristics like high vs.\ low power distance and agile challenges. 
The hypothetical correlation of an impact  is argued based on a detailed description of the analyzed data. However, the presented arguments per impact reference the data from Hofstede, which may be inaccurate for their case.
For example, the authors state~\cite{Ayed.2017}: \textit{``In BE teams, we observed a misunderstanding of team empowerment: T3 [...] reported that the team members were considered accountable for some business decisions and priorities definition. They also reported that they were sometimes confronted directly to the customer demands. The proximity of customer in itself is positive but this should not interfere on the team work during the iteration. The relatively high PDI score in BE can also explain the customer interference. 
In countries showing high PDI, hierarchy is well established and superiors, i.e., the customers in this case, may consider that they have special privileges such as asking the team for changes directly and anytime they want to."} 
The example shows, that the authors argue the observed findings based on Belgium Hofstede's score for the power distance dimension. 
However, if the team under study has a different power distance than Hofstede's data shows, the arguments for the observed behavior may not be correct and, thus,  different effects may be relevant. 

The second study is an interview study and investigates the motivation and challenges of an agile transition in Turkey~\cite{Gelmis.2022}.
The authors discuss their results and set relations to the data provided by Hofstede to argue potential cultural challenges and barriers in the Turkish context. For example, they state: \textit{``Thus, it is not surprising that study [...] report that Agile teams in the Asian countries suffer from high Power Distance and Uncertainty Avoidance as in Turkey. Similarly, as Asian countries, Turkey has high Power Distance, and Uncertainty Avoidance. Our results indicate that high Power Distance and Uncertainty Avoidance in Turkey result in negative impacts on factors such as team empowerment, feedback loops, securely failing, transparency, and process improvement."} Similar to the  study discussed above, we see a limitation related to the line of arguments, as the authors reference  the high PDI score in the data from Hofstede. Further, they generalize a potential indication on this cultural characteristic on specific aspects of agile methods. 

In accordance with the mentioned limitations in the papers, we verified also their citations using backward snowballing. It is seen with many concerns, that the study from Ayed et al.\ is already used for arguing similar cultural influences in other contexts, which is for example done by Gelmis et al.\ \cite{Gelmis.2022} (see the example above; Ayed et al.\ is one of the referenced studies in the cited statement shortened with [...]).

Based on our discussion, we derive three recommendations, which aim to increase the awareness of considering essential aspects when dealing with cultural influences:

\textit{Recommendation 1:} One should always be aware of the context under study~\cite{Dyba.2012}. As culture is multi-faceted, the behaviour of people is further influenced by many levels (in terms of regions, organizations, or sub-cultures). 
Further, people's behavior is shaped by the context, e.g., communities, in which they operate. 
Thus, we highly recommend to focus on at least two cultural levels, e.g., combining characteristics from organizational and regional/national cultures~\cite{Smite.2020,Smite.2021}. 

\textit{Recommendation 2:} Conducting studies dealing with the influences of cultural characteristics need a thorough validation of the used cultural values. 
We recommend to validate the cultural values especially based on qualitative data collection like observation or interviews, in order to be able to verify the self-determined cultural values or external provided data. Ethnographic study design may be a valuable approach for investigating studies in this context~\cite{Sharp.2016}. 

\textit{Recommendation 3:} In order to achieve a detailed understanding of the actual cultural situation in a case study environment, we propose to collect as much as possible the characteristics within the case context, ideally with the same people relevant to the study. 
This also directly leads to composed values resulting from the layering of the various cultural contexts (nation, region, corporate, project, etc.) the individuals are subject to. 

While we derived the above recommendations in the context of agile software development and this may have led to particularly pronounced deviations from the national average values measured by Hofstede, we believe that the recommendations above are relevant beyond this context. The reason is that also in other context, one cannot safely make assumptions about the cultural values without measuring them within the specific context.

\section{Conclusion}
\label{Sec6:Conclusion}
In this paper, we discussed challenges  in investigating cultural characteristics in the field of empirical software engineering, while using the well-known cultural dimensions model by Hofstede and focusing on agile software development. 

We conducted an exploratory multiple case study in four companies in Japan and Germany. The data was collected using the standardized questionnaire by Hofstede to be able to compare the determined cultural values of our participants with the scores from Hofstede. 
Our results show significant difference related to the data from Hofstede, in particular to dimensions uncertainty avoidance and long term orientation in Germany, and masculinity, uncertainty avoidance and long term orientation in Japan. 

Based on a thorough discussion of approaches used in existing research, we propose our contribution: Three recommendations, which should be taken into account in future research dealing with the influence of cultural characteristics, in order to increase the correctness and reliability of empirical research results in the context of culture. 


\bibliographystyle{ACM-Reference-Format}
\bibliography{references}


\begin{thebibliography}{54}


\ifx \showCODEN    \undefined \def \showCODEN     #1{\unskip}     \fi
\ifx \showDOI      \undefined \def \showDOI       #1{#1}\fi
\ifx \showISBNx    \undefined \def \showISBNx     #1{\unskip}     \fi
\ifx \showISBNxiii \undefined \def \showISBNxiii  #1{\unskip}     \fi
\ifx \showISSN     \undefined \def \showISSN      #1{\unskip}     \fi
\ifx \showLCCN     \undefined \def \showLCCN      #1{\unskip}     \fi
\ifx \shownote     \undefined \def \shownote      #1{#1}          \fi
\ifx \showarticletitle \undefined \def \showarticletitle #1{#1}   \fi
\ifx \showURL      \undefined \def \showURL       {\relax}        \fi
\providecommand\bibfield[2]{#2}
\providecommand\bibinfo[2]{#2}
\providecommand\natexlab[1]{#1}
\providecommand\showeprint[2][]{arXiv:#2}

\bibitem[Alsanoosy et~al\mbox{.}(2018a)]%
        {Alsanoosy.2018}
\bibfield{author}{\bibinfo{person}{T. Alsanoosy}, \bibinfo{person}{M. Spichkova}, {and} \bibinfo{person}{J. Harland}.} \bibinfo{year}{2018}\natexlab{a}.
\newblock \showarticletitle{Cultural Influences on Requirements Engineering Process in the Context of Saudi Arabia}. In \bibinfo{booktitle}{\emph{Proceedings of the 13th International Conference on Evaluation of Novel Approaches to Software Engineering}} (Funchal, Madeira, Portugal). \bibinfo{publisher}{SCITEPRESS - Science and Technology Publications, Lda}, \bibinfo{address}{Setubal, PRT}, \bibinfo{pages}{159–168}.
\newblock
\showISBNx{9789897583001}
\urldef\tempurl%
\url{https://doi.org/10.5220/0006770701590168}
\showDOI{\tempurl}


\bibitem[Alsanoosy et~al\mbox{.}(2018b)]%
        {Alsanoosy.2018a}
\bibfield{author}{\bibinfo{person}{T. Alsanoosy}, \bibinfo{person}{M. Spichkova}, {and} \bibinfo{person}{J. Harland}.} \bibinfo{year}{2018}\natexlab{b}.
\newblock \showarticletitle{Cultural Influences on the Requirements Engineering Process: Lessons Learned from Practice}. In \bibinfo{booktitle}{\emph{Proceedings of the 23rd International Conference on Engineering of Complex Computer Systems}}. \bibinfo{pages}{61--70}.
\newblock
\urldef\tempurl%
\url{https://doi.org/10.1109/ICECCS2018.2018.00015}
\showDOI{\tempurl}


\bibitem[Alsanoosy et~al\mbox{.}(2019)]%
        {Alsanoosy.2019}
\bibfield{author}{\bibinfo{person}{T. Alsanoosy}, \bibinfo{person}{M. Spichkova}, {and} \bibinfo{person}{J. Harland}.} \bibinfo{year}{2019}\natexlab{}.
\newblock \showarticletitle{The influence of power distance on requirements engineering activities}.
\newblock \bibinfo{journal}{\emph{Procedia Computer Science}}  \bibinfo{volume}{159} (\bibinfo{year}{2019}), \bibinfo{pages}{2394--2403}.
\newblock
\showISSN{1877-0509}
\urldef\tempurl%
\url{https://doi.org/10.1016/j.procs.2019.09.414}
\showDOI{\tempurl}


\bibitem[Alsanoosy et~al\mbox{.}(2020)]%
        {Alsanoosy.2020}
\bibfield{author}{\bibinfo{person}{T. Alsanoosy}, \bibinfo{person}{M. Spichkova}, {and} \bibinfo{person}{J. Harland}.} \bibinfo{year}{2020}\natexlab{}.
\newblock \showarticletitle{Cultural influence on requirements engineering activities: a systematic literature review and analysis}.
\newblock \bibinfo{journal}{\emph{Requirements Eng}}  \bibinfo{volume}{25} (\bibinfo{year}{2020}), \bibinfo{pages}{339–362}.
\newblock
\urldef\tempurl%
\url{https://doi.org/10.1007/s00766-019-00326-9}
\showDOI{\tempurl}


\bibitem[Amrit et~al\mbox{.}(2014)]%
        {Amrit.2014}
\bibfield{author}{\bibinfo{person}{C. Amrit}, \bibinfo{person}{M. Daneva}, {and} \bibinfo{person}{D. Damian}.} \bibinfo{year}{2014}\natexlab{}.
\newblock \showarticletitle{Human factors in software development: On its underlying theories and the value of learning from related disciplines. A guest editorial introduction to the special issue}.
\newblock \bibinfo{journal}{\emph{Information and Software Technology}} \bibinfo{volume}{56}, \bibinfo{number}{12} (\bibinfo{year}{2014}), \bibinfo{pages}{1537--1542}.
\newblock
\showISSN{09505849}


\bibitem[Ayed et~al\mbox{.}(2017)]%
        {Ayed.2017}
\bibfield{author}{\bibinfo{person}{H. Ayed}, \bibinfo{person}{B. Vanderose}, {and} \bibinfo{person}{N. Habra}.} \bibinfo{year}{2017}\natexlab{}.
\newblock \showarticletitle{Agile cultural challenges in {Europe} and {Asia}: insights from practitioners}. In \bibinfo{booktitle}{\emph{Intl. Conf. on Software Engineering: SEIP}}. \bibinfo{pages}{153--162}.
\newblock
\showISBNx{978-1-5386-2717-4}
\urldef\tempurl%
\url{https://doi.org/10.1109/ICSE-SEIP.2017.33}
\showDOI{\tempurl}


\bibitem[Baskerville(2003)]%
        {Baskerville.2003}
\bibfield{author}{\bibinfo{person}{R.~F. Baskerville}.} \bibinfo{year}{2003}\natexlab{}.
\newblock \showarticletitle{Hofstede never studied culture}.
\newblock \bibinfo{journal}{\emph{Accounting, Organizations and Society}} \bibinfo{volume}{28}, \bibinfo{number}{1} (\bibinfo{year}{2003}), \bibinfo{pages}{1--14}.
\newblock
\showISSN{0361-3682}
\urldef\tempurl%
\url{https://doi.org/10.1016/S0361-3682(01)00048-4}
\showDOI{\tempurl}


\bibitem[Beck(2000)]%
        {Beck.2000}
\bibfield{author}{\bibinfo{person}{K. Beck}.} \bibinfo{year}{2000}\natexlab{}.
\newblock \bibinfo{booktitle}{\emph{Extreme programming explained: Embrace change} (\bibinfo{edition}{5. print} ed.)}.
\newblock \bibinfo{publisher}{Addison-Wesley}, \bibinfo{address}{Boston}.
\newblock


\bibitem[Biddle et~al\mbox{.}(2018)]%
        {Biddle.2018}
\bibfield{author}{\bibinfo{person}{R. Biddle}, \bibinfo{person}{A. Meier}, \bibinfo{person}{M. Kropp}, {and} \bibinfo{person}{C. Anslow}.} \bibinfo{year}{2018}\natexlab{}.
\newblock \showarticletitle{Myagile: Sociological and Cultural Effects of Agile on Teams and Their Members}. In \bibinfo{booktitle}{\emph{Proceedings of the 11th International Workshop on Cooperative and Human Aspects of Software Engineering}}. \bibinfo{pages}{73–76}.
\newblock
\showISBNx{9781450357258}
\urldef\tempurl%
\url{https://doi.org/10.1145/3195836.3195845}
\showDOI{\tempurl}


\bibitem[Borchers(2003)]%
        {Borchers.2003}
\bibfield{author}{\bibinfo{person}{G. Borchers}.} \bibinfo{year}{2003}\natexlab{}.
\newblock \showarticletitle{The software engineering impacts of cultural factors on multi-cultural software development teams}. In \bibinfo{booktitle}{\emph{Proceedings of the 25th International Conference on Software Engineering}}. \bibinfo{pages}{540--545}.
\newblock
\urldef\tempurl%
\url{https://doi.org/10.1109/ICSE.2003.1201234}
\showDOI{\tempurl}


\bibitem[Brockmann and Thaumuller(2009)]%
        {Brockmann.2009}
\bibfield{author}{\bibinfo{person}{P.~S. Brockmann} {and} \bibinfo{person}{T. Thaumuller}.} \bibinfo{year}{2009}\natexlab{}.
\newblock \showarticletitle{Cultural Aspects of Global Requirements Engineering: An Empirical Chinese-German Case Study}. In \bibinfo{booktitle}{\emph{Proceedings of the 4th International Conference on Global Software Engineering}}. \bibinfo{pages}{353--357}.
\newblock
\urldef\tempurl%
\url{https://doi.org/10.1109/ICGSE.2009.55}
\showDOI{\tempurl}


\bibitem[Carmel and Agarwal(2001)]%
        {Carmel.2001}
\bibfield{author}{\bibinfo{person}{E. Carmel} {and} \bibinfo{person}{R. Agarwal}.} \bibinfo{year}{2001}\natexlab{}.
\newblock \showarticletitle{Tactical approaches for alleviating distance in global software development}.
\newblock \bibinfo{journal}{\emph{IEEE Software}} \bibinfo{volume}{18}, \bibinfo{number}{2} (\bibinfo{year}{2001}), \bibinfo{pages}{22--29}.
\newblock
\urldef\tempurl%
\url{https://doi.org/10.1109/52.914734}
\showDOI{\tempurl}


\bibitem[Chow and Cao(2008)]%
        {Chow.2008}
\bibfield{author}{\bibinfo{person}{T. Chow} {and} \bibinfo{person}{D.-B. Cao}.} \bibinfo{year}{2008}\natexlab{}.
\newblock \showarticletitle{A survey study of critical success factors in agile software projects}.
\newblock \bibinfo{journal}{\emph{Journal of Systems and Software}} \bibinfo{volume}{81}, \bibinfo{number}{6} (\bibinfo{year}{2008}), \bibinfo{pages}{961--971}.
\newblock
\showISSN{01641212}


\bibitem[Deshpande et~al\mbox{.}(2010)]%
        {Deshpande.2010}
\bibfield{author}{\bibinfo{person}{Sadhana Deshpande}, \bibinfo{person}{Ita Richardson}, \bibinfo{person}{Valentine Casey}, {and} \bibinfo{person}{Sarah Beecham}.} \bibinfo{year}{2010}\natexlab{}.
\newblock \showarticletitle{Culture in Global Software Development - A Weakness or Strength?}. In \bibinfo{booktitle}{\emph{Intl. Conf. on Global Software Engineering}}. \bibinfo{pages}{67--76}.
\newblock
\urldef\tempurl%
\url{https://doi.org/10.1109/ICGSE.2010.16}
\showDOI{\tempurl}


\bibitem[Diebold et~al\mbox{.}(2015)]%
        {Diebold.2015}
\bibfield{author}{\bibinfo{person}{P. Diebold}, \bibinfo{person}{S. Küpper}, {and} \bibinfo{person}{T. Zehler}.} \bibinfo{year}{2015}\natexlab{}.
\newblock \showarticletitle{Nachhaltige Agile Transition: Symbiose von technischer und kultureller Agilität}. In \bibinfo{booktitle}{\emph{Projektmanagement und Vorgehensmodelle}}. \bibinfo{pages}{121--126}.
\newblock


\bibitem[Dyb\r{a} et~al\mbox{.}(2012)]%
        {Dyba.2012}
\bibfield{author}{\bibinfo{person}{T. Dyb\r{a}}, \bibinfo{person}{D.~I.K. Sj\o{}berg}, {and} \bibinfo{person}{D.~S. Cruzes}.} \bibinfo{year}{2012}\natexlab{}.
\newblock \showarticletitle{What Works for Whom, Where, When, and Why? On the Role of Context in Empirical Software Engineering}. In \bibinfo{booktitle}{\emph{Proceedings of the ACM-IEEE International Symposium on Empirical Software Engineering and Measurement}}. \bibinfo{pages}{19–28}.
\newblock
\showISBNx{9781450310567}
\urldef\tempurl%
\url{https://doi.org/10.1145/2372251.2372256}
\showDOI{\tempurl}


\bibitem[Fischer et~al\mbox{.}(2021)]%
        {Fischer.2021}
\bibfield{author}{\bibinfo{person}{R.~A.-L. Fischer}, \bibinfo{person}{R. Walczuch}, {and} \bibinfo{person}{E. Guzman}.} \bibinfo{year}{2021}\natexlab{}.
\newblock \showarticletitle{Does Culture Matter? Impact of Individualism and Uncertainty Avoidance on App Reviews}. In \bibinfo{booktitle}{\emph{Proceedings of the 43rd International Conference on Software Engineering: Software Engineering in Society}}. \bibinfo{pages}{67--76}.
\newblock
\urldef\tempurl%
\url{https://doi.org/10.1109/ICSE-SEIS52602.2021.00016}
\showDOI{\tempurl}


\bibitem[Gelmis et~al\mbox{.}(2022)]%
        {Gelmis.2022}
\bibfield{author}{\bibinfo{person}{A. Gelmis}, \bibinfo{person}{N. Ozkan}, \bibinfo{person}{A.J. Ahmad}, {and} \bibinfo{person}{M.G. Guler}.} \bibinfo{year}{2022}\natexlab{}.
\newblock \showarticletitle{Impact of Turkish National Culture on Agile Software Development in Turkey}. In \bibinfo{booktitle}{\emph{Lean and Agile Software Development}}. \bibinfo{pages}{78--95}.
\newblock
\showISBNx{978-3-030-94238-0}


\bibitem[Ghinea et~al\mbox{.}(2011)]%
        {Ghinea.2011}
\bibfield{author}{\bibinfo{person}{G. Ghinea}, \bibinfo{person}{B. Bygstad}, {and} \bibinfo{person}{M. Satpathy}.} \bibinfo{year}{2011}\natexlab{}.
\newblock \showarticletitle{Software Developers in India and Norway: Professional or National Cultures?}
\newblock \bibinfo{journal}{\emph{Journal of Information Technology Research}} \bibinfo{volume}{4}, \bibinfo{number}{3} (\bibinfo{date}{jul} \bibinfo{year}{2011}), \bibinfo{pages}{50–63}.
\newblock
\showISSN{1938-7857}
\urldef\tempurl%
\url{https://doi.org/10.4018/jitr.2011070104}
\showDOI{\tempurl}


\bibitem[Gregory and Taylor(2019)]%
        {Gregory.2019}
\bibfield{author}{\bibinfo{person}{P. Gregory} {and} \bibinfo{person}{K. Taylor}.} \bibinfo{year}{2019}\natexlab{}.
\newblock \showarticletitle{Defining Agile Culture: A Collaborative and Practitioner-Led Approach}. In \bibinfo{booktitle}{\emph{Proceedings of the 12th International Workshop on Cooperative and Human Aspects of Software Engineering}}. \bibinfo{pages}{37--38}.
\newblock
\urldef\tempurl%
\url{https://doi.org/10.1109/CHASE.2019.00016}
\showDOI{\tempurl}


\bibitem[Gupta et~al\mbox{.}(2019)]%
        {Gupta.2019}
\bibfield{author}{\bibinfo{person}{M. Gupta}, \bibinfo{person}{J.F. George}, {and} \bibinfo{person}{W. Xia}.} \bibinfo{year}{2019}\natexlab{}.
\newblock \showarticletitle{Relationships between IT department culture and agile software development practices: An empirical investigation}.
\newblock \bibinfo{journal}{\emph{International Journal of Information Management}}  \bibinfo{volume}{44} (\bibinfo{year}{2019}), \bibinfo{pages}{13--24}.
\newblock
\showISSN{0268-4012}
\urldef\tempurl%
\url{https://doi.org/10.1016/j.ijinfomgt.2018.09.006}
\showDOI{\tempurl}


\bibitem[Guzman et~al\mbox{.}(2018)]%
        {Guzman.2018}
\bibfield{author}{\bibinfo{person}{E. Guzman}, \bibinfo{person}{L. Oliveira}, \bibinfo{person}{Y. Steiner}, \bibinfo{person}{L.~C. Wagner}, {and} \bibinfo{person}{M. Glinz}.} \bibinfo{year}{2018}\natexlab{}.
\newblock \showarticletitle{User feedback in the app store: a cross-cultural study}. In \bibinfo{booktitle}{\emph{Proceedings of the 40th International Conference on Software Engineering: Software Engineering in Society}} (Gothenburg, Sweden). \bibinfo{publisher}{Association for Computing Machinery}, \bibinfo{address}{New York, NY, USA}, \bibinfo{pages}{13–22}.
\newblock
\showISBNx{9781450356619}
\urldef\tempurl%
\url{https://doi.org/10.1145/3183428.3183436}
\showDOI{\tempurl}


\bibitem[Hoda et~al\mbox{.}(2017)]%
        {Hoda.2017}
\bibfield{author}{\bibinfo{person}{R. Hoda}, \bibinfo{person}{N. Salleh}, \bibinfo{person}{J. Grundy}, {and} \bibinfo{person}{H.~M. Tee}.} \bibinfo{year}{2017}\natexlab{}.
\newblock \showarticletitle{Systematic literature reviews in agile software development: A tertiary study}.
\newblock \bibinfo{journal}{\emph{Information and Software Technology}}  \bibinfo{volume}{85} (\bibinfo{year}{2017}), \bibinfo{pages}{60--70}.
\newblock
\showISSN{0950-5849}
\urldef\tempurl%
\url{https://doi.org/10.1016/j.infsof.2017.01.007}
\showDOI{\tempurl}


\bibitem[Hofstede(1981)]%
        {Hofstede.1981}
\bibfield{author}{\bibinfo{person}{G. Hofstede}.} \bibinfo{year}{1981}\natexlab{}.
\newblock \bibinfo{booktitle}{\emph{Culture's consequences: international differences in work-related values}}.
\newblock \bibinfo{publisher}{{Sage Publications}}.
\newblock


\bibitem[Hofstede(2001)]%
        {Hofstede.2001}
\bibfield{author}{\bibinfo{person}{G. Hofstede}.} \bibinfo{year}{2001}\natexlab{}.
\newblock \bibinfo{booktitle}{\emph{Culture's consequences: Comparing values, behaviors, institutions, and organizations across nations} (\bibinfo{edition}{2nd} ed.)}.
\newblock \bibinfo{publisher}{{Sage Publications}}.
\newblock
\showISBNx{9780803973244}


\bibitem[Hofstede(2010)]%
        {HofstedeWebsite.2010}
\bibfield{author}{\bibinfo{person}{G. Hofstede}.} \bibinfo{year}{2010}\natexlab{}.
\newblock \bibinfo{title}{Hofstede Cultural Dimensions - Country Comparison Graphs}.
\newblock
\newblock
\newblock
\shownote{https://geerthofstede.com/country-comparison-graphs/}.


\bibitem[Hofstede et~al\mbox{.}(2010)]%
        {Hofstede.2010}
\bibfield{author}{\bibinfo{person}{G. Hofstede}, \bibinfo{person}{G.~J. Hofstede}, {and} \bibinfo{person}{M. Minkov}.} \bibinfo{year}{2010}\natexlab{}.
\newblock \bibinfo{booktitle}{\emph{Cultures and organizations. Software of the mind ; intercultural co-operation and its importance for survival} (\bibinfo{edition}{3rd} ed.)}.
\newblock \bibinfo{publisher}{{McGraw-Hill Education}}.
\newblock


\bibitem[Hofstede and Minkov(2013)]%
        {HofstedeMinkov.2013}
\bibfield{author}{\bibinfo{person}{G. Hofstede} {and} \bibinfo{person}{M. Minkov}.} \bibinfo{year}{2013}\natexlab{}.
\newblock \bibinfo{title}{Values Survey Module Manual}.
\newblock
\newblock
\newblock
\shownote{https://geerthofstede.com/wp-content/uploads/2016/07/Manual-VSM-2013.pdf}.


\bibitem[House et~al\mbox{.}(2002)]%
        {House.2002}
\bibfield{author}{\bibinfo{person}{R. House}, \bibinfo{person}{M. Javidan}, \bibinfo{person}{P. Hanges}, {and} \bibinfo{person}{P. Dorfman}.} \bibinfo{year}{2002}\natexlab{}.
\newblock \showarticletitle{Understanding cultures and implicit leadership theories across the globe: an introduction to project GLOBE}.
\newblock \bibinfo{journal}{\emph{Journal of World Business}} \bibinfo{volume}{37}, \bibinfo{number}{1} (\bibinfo{year}{2002}), \bibinfo{pages}{3--10}.
\newblock
\showISSN{10909516}
\urldef\tempurl%
\url{https://doi.org/10.1016/S1090-9516(01)00069-4}
\showDOI{\tempurl}


\bibitem[Hwang(2012)]%
        {Hwang.2012}
\bibfield{author}{\bibinfo{person}{Y. Hwang}.} \bibinfo{year}{2012}\natexlab{}.
\newblock \showarticletitle{End User Adoption of Enterprise Systems in Eastern and Western Cultures.}
\newblock \bibinfo{journal}{\emph{Journal of Organizational and End User Computing}} \bibinfo{volume}{24}, \bibinfo{number}{4} (\bibinfo{year}{2012}), \bibinfo{pages}{1--17}.
\newblock
\urldef\tempurl%
\url{https://doi.org/10.4018/joeuc.2012100101}
\showDOI{\tempurl}


\bibitem[Iivari and Iivari(2011)]%
        {Iivari.2011}
\bibfield{author}{\bibinfo{person}{J. Iivari} {and} \bibinfo{person}{N. Iivari}.} \bibinfo{year}{2011}\natexlab{}.
\newblock \showarticletitle{The relationship between organizational culture and the deployment of agile methods}.
\newblock \bibinfo{journal}{\emph{Information and Software Technology}} \bibinfo{volume}{53}, \bibinfo{number}{5} (\bibinfo{year}{2011}), \bibinfo{pages}{509--520}.
\newblock
\showISSN{0950-5849}
\urldef\tempurl%
\url{https://doi.org/10.1016/j.infsof.2010.10.008}
\showDOI{\tempurl}


\bibitem[Insights(2024)]%
        {HofstedeInsights.2024}
\bibfield{author}{\bibinfo{person}{Hofstede Insights}.} \bibinfo{year}{2024}\natexlab{}.
\newblock \bibinfo{title}{Country comparison}.
\newblock
\newblock
\newblock
\shownote{https://www.hofstede-insights.com/country-comparison/}.


\bibitem[ITIM(2017)]%
        {Cultural_Survey}
\bibfield{author}{\bibinfo{person}{ITIM}.} \bibinfo{year}{2017}\natexlab{}.
\newblock \bibinfo{booktitle}{\emph{{A Cultural Survey}}}.
\newblock
\urldef\tempurl%
\url{https://doi.org/10.5281/zenodo.10581893}
\showDOI{\tempurl}


\bibitem[Jaanu et~al\mbox{.}(2012)]%
        {Jaanu.2012}
\bibfield{author}{\bibinfo{person}{T. Jaanu}, \bibinfo{person}{M. Paasivaara}, {and} \bibinfo{person}{C. Lassenius}.} \bibinfo{year}{2012}\natexlab{}.
\newblock \showarticletitle{Effects of four distances on communication processes in global software projects}. In \bibinfo{booktitle}{\emph{Proceedings of the ACM-IEEE International Symposium on Empirical Software Engineering and Measurement}}. \bibinfo{pages}{231--234}.
\newblock
\urldef\tempurl%
\url{https://doi.org/10.1145/2372251.2372293}
\showDOI{\tempurl}


\bibitem[Karahanna et~al\mbox{.}(2006)]%
        {Karahanna.2006}
\bibfield{author}{\bibinfo{person}{E. Karahanna}, \bibinfo{person}{J.~R. Evaristo}, {and} \bibinfo{person}{M. Srite}.} \bibinfo{year}{2006}\natexlab{}.
\newblock \showarticletitle{Levels of Culture and Individual Behavior: An Integrative Perspective}.
\newblock \bibinfo{journal}{\emph{Adv. topics in global information management}} (\bibinfo{year}{2006}), \bibinfo{pages}{30--50}.
\newblock
\showISBNx{9781591409236}
\urldef\tempurl%
\url{https://doi.org/10.4018/978-1-59140-923-6.ch002}
\showDOI{\tempurl}


\bibitem[Kuchel et~al\mbox{.}(2023)]%
        {Kuchel.2023}
\bibfield{author}{\bibinfo{person}{T. Kuchel}, \bibinfo{person}{M. Neumann}, \bibinfo{person}{P. Diebold}, {and} \bibinfo{person}{E.-M. Sch\"{o}n}.} \bibinfo{year}{2023}\natexlab{}.
\newblock \showarticletitle{Which Challenges Do Exist With Agile Culture in Practice?}. In \bibinfo{booktitle}{\emph{Proceedings of the 38th ACM/SIGAPP Symposium on Applied Computing}} (Tallinn, Estonia) \emph{(\bibinfo{series}{SAC '23})}. \bibinfo{publisher}{Association for Computing Machinery}, \bibinfo{address}{New York, NY, USA}, \bibinfo{pages}{1018–1025}.
\newblock
\showISBNx{9781450395175}
\urldef\tempurl%
\url{https://doi.org/10.1145/3555776.3578726}
\showDOI{\tempurl}


\bibitem[Küpper et~al\mbox{.}(2017)]%
        {Kuepper.2017}
\bibfield{author}{\bibinfo{person}{S. Küpper}, \bibinfo{person}{M. Kuhrmann}, \bibinfo{person}{M. Wiatrok}, \bibinfo{person}{U. Andelfinger}, {and} \bibinfo{person}{A. Rausch}.} \bibinfo{year}{2017}\natexlab{}.
\newblock \showarticletitle{Is There a Blueprint for Building an Agile Culture?}. In \bibinfo{booktitle}{\emph{Projektmanagement und Vorgehensmodelle}}. \bibinfo{pages}{111--128}.
\newblock


\bibitem[Martinsons and Davison(2007)]%
        {Maris.2007}
\bibfield{author}{\bibinfo{person}{M.~G. Martinsons} {and} \bibinfo{person}{R.~M. Davison}.} \bibinfo{year}{2007}\natexlab{}.
\newblock \showarticletitle{Strategic decision making and support systems: Comparing American, Japanese and Chinese management}.
\newblock \bibinfo{journal}{\emph{Decision Support Systems}} \bibinfo{volume}{43}, \bibinfo{number}{1} (\bibinfo{year}{2007}), \bibinfo{pages}{284--300}.
\newblock
\showISSN{0167-9236}
\urldef\tempurl%
\url{https://doi.org/10.1016/j.dss.2006.10.005}
\showDOI{\tempurl}
\newblock
\shownote{Mobile Commerce: Strategies, Technologies, and Applications}.


\bibitem[McSweeney(2002)]%
        {McSweeney.2002}
\bibfield{author}{\bibinfo{person}{B. McSweeney}.} \bibinfo{year}{2002}\natexlab{}.
\newblock \showarticletitle{Hofstede’s Model of National Cultural Differences and their Consequences: A Triumph of Faith - a Failure of Analysis}.
\newblock \bibinfo{journal}{\emph{Human Relations}} \bibinfo{volume}{55}, \bibinfo{number}{1} (\bibinfo{year}{2002}), \bibinfo{pages}{89--118}.
\newblock
\urldef\tempurl%
\url{https://doi.org/10.1177/0018726702551004}
\showDOI{\tempurl}


\bibitem[Neumann(2024)]%
        {Dataset_CulturalProfiles}
\bibfield{author}{\bibinfo{person}{M. Neumann}.} \bibinfo{year}{2024}\natexlab{}.
\newblock \bibinfo{booktitle}{\emph{{Dataset: Cultural Profiles for All Cases}}}.
\newblock
\urldef\tempurl%
\url{https://doi.org/10.5281/zenodo.10800926}
\showDOI{\tempurl}


\bibitem[Neumann et~al\mbox{.}(2023)]%
        {Neumann.2023}
\bibfield{author}{\bibinfo{person}{M. Neumann}, \bibinfo{person}{K. Schmid}, {and} \bibinfo{person}{L. Baumann}.} \bibinfo{year}{2023}\natexlab{}.
\newblock \showarticletitle{Characterizing The Impact of Culture on Agile Methods: The MoCA Model}. In \bibinfo{booktitle}{\emph{Proceedings of the 17th International Conference on Software and System Processes}}. \bibinfo{publisher}{IEEE}, \bibinfo{pages}{81--85}.
\newblock
\urldef\tempurl%
\url{https://doi.org/10.1109/ICSSP59042.2023.00018}
\showDOI{\tempurl}


\bibitem[Rabelo et~al\mbox{.}(2015)]%
        {Rabelo.2015}
\bibfield{author}{\bibinfo{person}{J.~d.~H. Rabelo}, \bibinfo{person}{E.~C. C.~d. Oliveira}, \bibinfo{person}{D. Viana}, \bibinfo{person}{L.~C. d.~S. Braga}, \bibinfo{person}{G.~d.~S. Souza}, \bibinfo{person}{I.~F. Steinmacher}, {and} \bibinfo{person}{T.~U. Conte}.} \bibinfo{year}{2015}\natexlab{}.
\newblock \showarticletitle{Knowledge Management and Organizational Culture in a Software Organization -- A Case Study}. In \bibinfo{booktitle}{\emph{Proceedings of the 8th International Workshop on Cooperative and Human Aspects of Software Engineering}}. \bibinfo{pages}{89--92}.
\newblock
\urldef\tempurl%
\url{https://doi.org/10.1109/CHASE.2015.27}
\showDOI{\tempurl}


\bibitem[Runeson and H{\"o}st(2009)]%
        {Runeson.2009}
\bibfield{author}{\bibinfo{person}{P. Runeson} {and} \bibinfo{person}{M. H{\"o}st}.} \bibinfo{year}{2009}\natexlab{}.
\newblock \showarticletitle{Guidelines for conducting and reporting case study research in software engineering}.
\newblock \bibinfo{journal}{\emph{Empirical Software Engineering}} \bibinfo{volume}{14}, \bibinfo{number}{2} (\bibinfo{year}{2009}), \bibinfo{pages}{131--164}.
\newblock
\showISSN{1382-3256}
\urldef\tempurl%
\url{https://doi.org/10.1007/s10664-008-9102-8}
\showDOI{\tempurl}


\bibitem[Schmitz and Weber(2014)]%
        {Schmitz.2014}
\bibfield{author}{\bibinfo{person}{L. Schmitz} {and} \bibinfo{person}{W. Weber}.} \bibinfo{year}{2014}\natexlab{}.
\newblock \showarticletitle{Are Hofstede's dimensions valid? A test for measurement invariance of uncertainty avoidance}.
\newblock \bibinfo{journal}{\emph{interculture journal: Online-Zeitschrift für interkulturelle Studien}} \bibinfo{volume}{13}, \bibinfo{number}{22} (\bibinfo{year}{2014}), \bibinfo{pages}{11--26}.
\newblock
\showISSN{2196-9485}


\bibitem[Schwaber and Sutherland(2020)]%
        {Schwaber.2020}
\bibfield{author}{\bibinfo{person}{K. Schwaber} {and} \bibinfo{person}{J. Sutherland}.} \bibinfo{year}{2020}\natexlab{}.
\newblock \bibinfo{title}{The Scrum Guide}.
\newblock
\newblock
\urldef\tempurl%
\url{https://www.scrumguides.org/scrum-guide.html}
\showURL{%
\tempurl}


\bibitem[Sharp et~al\mbox{.}(2016)]%
        {Sharp.2016}
\bibfield{author}{\bibinfo{person}{H. Sharp}, \bibinfo{person}{Y. Dittrich}, {and} \bibinfo{person}{C.~R.~B. de Souza}.} \bibinfo{year}{2016}\natexlab{}.
\newblock \showarticletitle{The Role of Ethnographic Studies in Empirical Software Engineering}.
\newblock \bibinfo{journal}{\emph{IEEE Transactions on Software Engineering}} \bibinfo{volume}{42}, \bibinfo{number}{8} (\bibinfo{year}{2016}), \bibinfo{pages}{786--804}.
\newblock
\urldef\tempurl%
\url{https://doi.org/10.1109/TSE.2016.2519887}
\showDOI{\tempurl}


\bibitem[Sharp and Robinson(2005)]%
        {Sharp.2005}
\bibfield{author}{\bibinfo{person}{H. Sharp} {and} \bibinfo{person}{H. Robinson}.} \bibinfo{year}{2005}\natexlab{}.
\newblock \showarticletitle{Some Social Factors of Software Engineering: The Maverick, Community and Technical Practices}. In \bibinfo{booktitle}{\emph{Proceedings of the Workshop on Human and Social Factors of Software Engineering}} \emph{(\bibinfo{series}{HSSE '05})}. \bibinfo{pages}{1–6}.
\newblock
\showISBNx{1595931201}
\urldef\tempurl%
\url{https://doi.org/10.1145/1083106.1083117}
\showDOI{\tempurl}


\bibitem[Sharp et~al\mbox{.}(1999)]%
        {Sharp.1999}
\bibfield{author}{\bibinfo{person}{H. Sharp}, \bibinfo{person}{M. Woodman}, \bibinfo{person}{F. Hovenden}, {and} \bibinfo{person}{H. Robinson}.} \bibinfo{year}{1999}\natexlab{}.
\newblock \showarticletitle{The role of 'culture' in successful software process improvement}. In \bibinfo{booktitle}{\emph{Proceedings 25th EUROMICRO Conference. Informatics: Theory and Practice for the New Millennium}}, Vol.~\bibinfo{volume}{2}. \bibinfo{pages}{170--176 vol.2}.
\newblock
\urldef\tempurl%
\url{https://doi.org/10.1109/EURMIC.1999.794778}
\showDOI{\tempurl}


\bibitem[Silveira and Prikladnicki(2019)]%
        {Silveira.2019}
\bibfield{author}{\bibinfo{person}{K.~K. Silveira} {and} \bibinfo{person}{R. Prikladnicki}.} \bibinfo{year}{2019}\natexlab{}.
\newblock \showarticletitle{A Systematic Mapping Study of Diversity in Software Engineering: A Perspective from the Agile Methodologies}. In \bibinfo{booktitle}{\emph{Proceedings of the 12th International Workshop on Cooperative and Human Aspects of Software Engineering}}. \bibinfo{pages}{7–10}.
\newblock
\urldef\tempurl%
\url{https://doi.org/10.1109/CHASE.2019.00010}
\showDOI{\tempurl}


\bibitem[{\v{S}}mite et~al\mbox{.}(2020)]%
        {Smite.2020}
\bibfield{author}{\bibinfo{person}{D. {\v{S}}mite}, \bibinfo{person}{J. Gonzalez-Huerta}, {and} \bibinfo{person}{N.~B. Moe}.} \bibinfo{year}{2020}\natexlab{}.
\newblock \showarticletitle{``When in Rome, Do as the Romans Do'': Cultural Barriers to Being Agile in Distributed Teams}. In \bibinfo{booktitle}{\emph{Agile Processes in Software Engineering and Extreme Programming}}. \bibinfo{pages}{145--161}.
\newblock
\showISBNx{978-3-030-49392-9}


\bibitem[Strode et~al\mbox{.}(2022)]%
        {Strode.2022}
\bibfield{author}{\bibinfo{person}{D.~E. Strode}, \bibinfo{person}{H. Sharp}, \bibinfo{person}{L. Barroca}, \bibinfo{person}{P. Gregory}, {and} \bibinfo{person}{K. Taylor}.} \bibinfo{year}{2022}\natexlab{}.
\newblock \showarticletitle{Tensions in Organizations Transforming to Agility}.
\newblock \bibinfo{journal}{\emph{IEEE Transactions on Engineering Management}} \bibinfo{volume}{69}, \bibinfo{number}{6} (\bibinfo{year}{2022}), \bibinfo{pages}{3572--3583}.
\newblock
\urldef\tempurl%
\url{https://doi.org/10.1109/TEM.2022.3160415}
\showDOI{\tempurl}


\bibitem[Trompenaars and Hampden-Turner(2012)]%
        {Trompenaars.2012}
\bibfield{author}{\bibinfo{person}{F. Trompenaars} {and} \bibinfo{person}{C. Hampden-Turner}.} \bibinfo{year}{2012}\natexlab{}.
\newblock \bibinfo{booktitle}{\emph{Riding the waves of culture: Understanding diversity in global business}}.
\newblock \bibinfo{publisher}{McGraw-Hill}.
\newblock
\showISBNx{9781904838388}


\bibitem[VersionOne and Collabnet(2022)]%
        {VersionOne.2022}
\bibfield{author}{\bibinfo{person}{VersionOne} {and} \bibinfo{person}{Collabnet}.} \bibinfo{year}{2022}\natexlab{}.
\newblock \bibinfo{title}{16th Annual State of Agile Survey}.
\newblock
\newblock
\newblock
\shownote{stateofagile.com}.


\bibitem[Šmite et~al\mbox{.}(2021)]%
        {Smite.2021}
\bibfield{author}{\bibinfo{person}{D. Šmite}, \bibinfo{person}{N.~B. Moe}, {and} \bibinfo{person}{J. Gonzalez-Huerta}.} \bibinfo{year}{2021}\natexlab{}.
\newblock \showarticletitle{Overcoming cultural barriers to being agile in distributed teams}.
\newblock \bibinfo{journal}{\emph{Information and Software Technology}}  \bibinfo{volume}{138} (\bibinfo{year}{2021}), \bibinfo{pages}{106612}.
\newblock
\showISSN{0950-5849}
\urldef\tempurl%
\url{https://doi.org/10.1016/j.infsof.2021.106612}
\showDOI{\tempurl}


\end{thebibliography}





\end{document}